\documentclass[preprint,12pt,review]{elsarticle}

\usepackage{amssymb}
\usepackage{amsmath}

\journal{arXiv}

\begin{document}

\begin{frontmatter}



\title{Particle decays in the presence of a neutrino background}


\author{I. Alikhanov\corref{cor1}}

\cortext[cor1]{{\it Email address: {\tt ialspbu@gmail.com}}}

\address{Institute for Nuclear Research of the Russian Academy of Sciences,
60-th October Anniversary pr. 7a, Moscow 117312, Russia}

\begin{abstract}
Several non-threshold reactions which may be used to detect the cosmic neutrino background are presented. The corresponding cross sections are calculated analytically within the Standard Model. These reactions are sensitive not only to the electron neutrinos  but also to the muon and tau neutrinos. Some possibilities of experimental observation of the neutrino background are indicated.
\end{abstract}


\end{frontmatter}


\section{Introduction}
The standard Big Bang cosmology predicts, along with the cosmic microwave background, the existence of a cosmic neutrino background (CNB). Its present temperature (the mean kinetic energy of the neutrinos) is very low -- about~1.95~K, which corresponds to the neutrino number density around~56~cm$^{-3}$ for each active neutrino species.

Presently the existence  of the CNB is suggested only by cosmological measurements~\cite{indirect}.
A direct observation of the CNB is still a challenging problem. Various strategies have been proposed for searches of the relic neutrinos in laboratories~\cite{revm1,revm2,ringwald09}. A perspective one is the capture of the CNB neutrinos on beta decaying nuclei~\cite{ringwald09,rev1,rev2,rev3,rev5,rev6,rev7,rev4}.

 Several additional weak interaction based possibilities of detection of the CNB are presented in this Letter.

\section{Induced muon decay}

The muon decay
\begin{equation}
\mu^+\rightarrow e^++\nu_e+\bar\nu_{\mu}
\label{muon_d}
\end{equation}

is very well known. It has entered textbooks many years ago and already become an integral part of every course on weak interactions.

Let us consider the following reaction related to (\ref{muon_d}) by crossing symmetry:

\begin{equation}
\nu_{\mu}+\mu^+\rightarrow e^++\nu_e.
\label{muon_dm}
\end{equation}

A notable property of the reaction (\ref{muon_dm}) is that it has  no energy threshold being therefore able to proceed even when the incident neutrino three-momentum $p_{\nu_{\mu}}$ tends to zero. This is crucial for detecting the CNB. The present temperature of the CNB predicted by standard Big Bang cosmology is about $10^{-4}$ eV which implies that at least two of the relic neutrino mass eigenstates may be non-relativistic today, i.e.
$m_{\nu_i}\gg p_{{\nu}_i}$. It is interesting to note that in an alternative world in which physicists have developed the Standard Model but have not yet predicted the existence of the CNB, an observation of (\ref{muon_dm}) would be thought of as a lepton number violating decay mode of the muon.

It is possible to distinguish (\ref{muon_d}) from (\ref{muon_dm}) provided the energy resolution of the detector is less than $m_{\nu_{\mu}}/2$.  Actually, due to the non-zero neutrino mass there is a gap between the maximum energies of the positrons emitted in these reactions given by

\begin{equation}
\Delta E=\frac{m_{\nu_{\mu}}}{2}\left[\frac{(m_{\mu}+m_{\nu_{\mu}}+m_{\nu_e})^2-m_e^2}{m_{\mu}(m_{\mu}+m_{\nu_{\mu}})}\right]\approx\frac{m_{\nu_{\mu}}}{2}.
\label{deltae}
\end{equation}

\begin{figure}
\centering
\resizebox{0.9\textwidth}{!}{%
\includegraphics{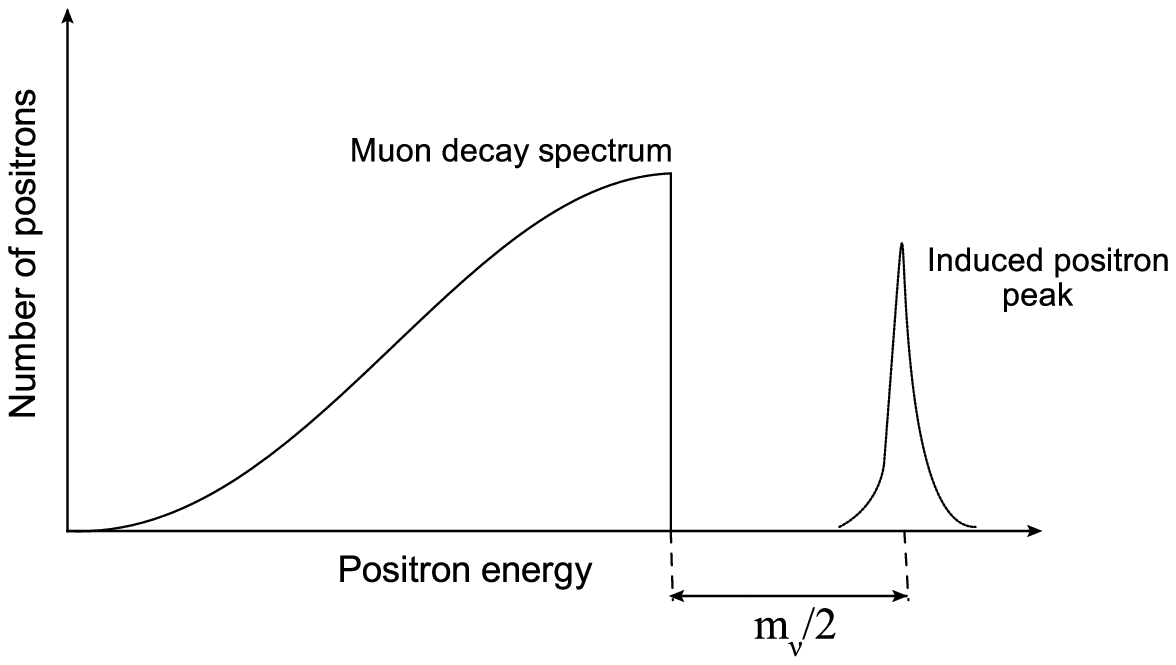}
}\caption{Qualitative illustration of the spectrum of positrons emitted in the process $\mu^+\rightarrow~e^+\nu_e\bar\nu_{\mu}$ plus $\nu_{\mu}\mu^+\rightarrow e^+\nu_e$.}
\label{fig1}
\end{figure}

This situation is qualitatively illustrated in Fig.~\ref{fig1}.

\begin{figure}
\centering
\resizebox{0.3\textwidth}{!}{%
\includegraphics{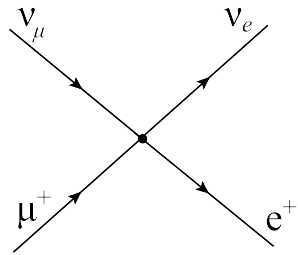}
}\caption{Feynman diagram for the process $\nu_{\mu}\mu^+\rightarrow e^+\nu_e$.}
\label{fig2}
\end{figure}

Since the momentum transfer in (\ref{muon_dm}) at such conditions is much smaller than the $W$ boson mass, it is enough to use the effective Fermi interaction to calculate the cross section. Then the corresponding matrix element represented by the Feynman diagram in Fig.~\ref{fig2} is

\begin{equation}
{\cal M}=-i\frac{G_F}{\sqrt{2}}\bar v_{\mu}\gamma^{\alpha}(1-\gamma^5)u_{\nu_{\mu}}\bar u_{\nu_e}\gamma_{\alpha}(1-\gamma^5)v_{e}
\label{muon_amplitude}
\end{equation}

so that the cross section for (\ref{muon_dm}) at $p_{\nu_{\mu}}\rightarrow0$ reads

\begin{equation}
\sigma_{\mu}=\frac{G^2_Fm^2_{\mu}}{2\pi}\left(1+\frac{m^2_e}{m^2_{\mu}}\right)\left(1-\frac{m^2_e}{m^2_{\mu}}\right)^2.
\label{muon_cross}
\end{equation}

Here $G_F$ is the Fermi coupling constant. All the neutrino masses are neglected in (\ref{muon_cross}).
The contribution of the terms $\sim m^2_e$ to the cross section is also negligible for the purposes of this Letter and can be safely omitted.
This will be done in the subsequent analysis.

One can find the mean lifetime of the muon at rest with respect to the "decay"~(\ref{muon_dm}) as

\begin{equation}
\bar\tau_{\mu}(\nu_{\mu})=\frac{1}{\sigma_{\mu}j_{\nu_{\mu}}},
\label{muon_life}
\end{equation}

where $j_{\nu_{\mu}}$ is the flux of the CNB neutrinos in the muon rest frame given by

\begin{equation}
j_{\nu_{\mu}}=n_{\nu_{\mu}}v.
\label{neutr_flux}
\end{equation}

Here $n_{\nu_{\mu}}$ is the number density of the CNB neutrinos, $v$ denotes the mean velocity of the relic neutrinos in the rest frame.
An estimation of the present-day mean lifetime in an earth-based laboratory ($v\approx7.7\times10^{-4}c$) leads to a large value

\begin{equation}
\bar\tau_{\mu}(\nu_{\mu})\approx2.7\times10^{23}\,\, \text{yr}\,\left(\frac{9.06\times10^{-41} \text{cm}^2}{\sigma_{\mu}}\right)\left(\frac{56~\text{cm}^{-3}}{n_{\nu_{\mu}}}\right)\left(\frac{7.7\times10^{-4}c}{v}\right).
\label{muon_life2}
\end{equation}

The muon neutrino content of the CNB can be probed via (\ref{muon_dm}). However, one should take into consideration that the muon lifetime is obviously tiny in comparison to a duration of any reasonable experiment. In general, for an unstable target particle, one should take the time average of the sample accounting for the decay: $\langle{N}\rangle=1/T\int_0^T N_0e^{-t/\tau}dt$, where $T$ is the exposure time of the experiment, $N_0$ is the number of target particles, $\tau$ is the target particle lifetime. So, one has that $\langle{N}\rangle\sim N_0(\tau/T)$. For example, to observe several events in the case of muons, for which $\tau=2.2\times10^{-6}$ s, in a year time scale experiment ($T\sim1$ year), one finds using~(\ref{muon_life2}) that one needs $N_0\sim10^{36}$ muons. This number is about $10^{12}$ times larger than that anticipated in the recently proposed Mu2e experiment at Fermilab~\cite{mu2e} which will produce~$10^{22}$ muons during its initial two-year running period. A second-phase, upgraded Mu2e experiment could, utilizing Fermilab's proposed Project X, a high-intensity proton accelerator, increase the production of muons by two orders of magnitude to roughly~$10^{24}$. The Mu2e could detect the induced muon decay, if gravitational clustering of the relic neutrinos~\cite{cluster} compensating the factor ($\tau/T$) takes place. A deviation from the Standard Model prediction for the cross section $\sigma_{\mu}$ is also possible.

There is another interesting perspective of detection of the CNB related to astrophysics. Calculations show that muons may appear inside compact stars, and at the densities of matter  of a few times nuclear state densities the muon density becomes of the order of $10^{37}$ cm$^{-3}$~\cite{comp_star}. These muons may feel the CNB as well and decaying via (\ref{muon_dm}) may affect the cooling processes of the stars. Detailed calculations are needed here and this is a subject of a separate consideration.

One also finds the cross section for an additional non-threshold reaction induced by the electron neutrino content of the CNB

\begin{equation}
\bar\nu_{e}+\mu^+\rightarrow e^++\bar\nu_{\mu}
\label{muon_dm2}
\end{equation}

to be

\begin{equation}
\sigma_{e}=\frac{G^2_Fm^2_{\mu}}{\pi}.
\label{muon_cross2}
\end{equation}

The corresponding mean lifetime of the muon is then determined by

\begin{equation}
\bar\tau_{\mu}(\bar\nu_{e})=\frac{\bar\tau_{\mu}(\nu_{\mu})}{2}
\label{muon_life3}
\end{equation}

provided $n_{\bar\nu_{e}}=n_{\nu_{\mu}}$.

Discussion and conclusions regarding the reaction~(\ref{muon_dm2}) are analogous to the ones given above. The formalism of this Letter directly applies to the tau lepton as well. All the characteristics of the reactions presented in this section will coincide with those of their CP conjugates ($\bar\nu_{\mu}+\mu^-\rightarrow e^-+\bar\nu_e$, etc.) if CP is conserved.

\section{Induced neutral pion decay}
The Standard Model accommodates the following reaction~\cite{Harvey2007,plb_pion}:

\begin{equation}
\nu_{l}+\pi^0\rightarrow\nu_{l}+\gamma
\label{pion_d0}
\end{equation}

which also has no threshold ($l=e, \mu, \tau$).

\begin{figure}
\centering
\resizebox{0.3\textwidth}{!}{%
\includegraphics{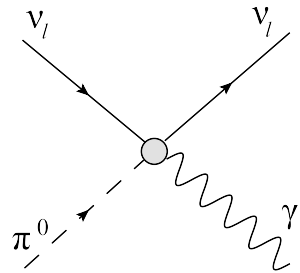}
}\caption{Feynman diagram for the process $\nu_{l}\pi^0\rightarrow\nu_{l}\gamma$.}
\label{fig3}
\end{figure}

The corresponding matrix element represented by the Feynman diagram in Fig.~\ref{fig3} is~\cite{plb_pion}

\begin{equation}
{\cal M}=\frac{eG_{F}}{\sqrt{2}m_{\pi}}F_{V}\varepsilon_{\mu }
\bar{u}(p'_{\nu})\gamma _{\alpha }(1-\gamma
_{5})u(p_{\nu})
\epsilon ^{\mu \alpha \beta
\lambda }q_{\beta }p_{\pi \lambda }. \label{ampl_2}
\end{equation}

Here $e$ is the elementary electric charge, $\varepsilon_{\mu}$ denotes the photon polarization vector,
$p_{\pi}$, $p_{\nu}$, $p'_{\nu}$, and $q$
are the four-momenta of $\pi^{0}$, the initial and final neutrinos and $\gamma$, respectively, $F_{V}$ is
the vector form factor.

After the standard algebra one obtains the cross section of~(\ref{pion_d0}) for each neutrino flavor in the limit $p_{\nu_l}\rightarrow0$:

\begin{equation}
\sigma_{\pi}=\frac{\alpha G^2_{F}m^2_\pi}{8}|F_{V}|^2, \label{cross_sect_tot2}
\end{equation}

where $\alpha$ is the fine structure constant.
Details of similar calculations can be found in~\cite{plb_pion}.

\begin{figure}
\centering
\resizebox{0.9\textwidth}{!}{%
\includegraphics{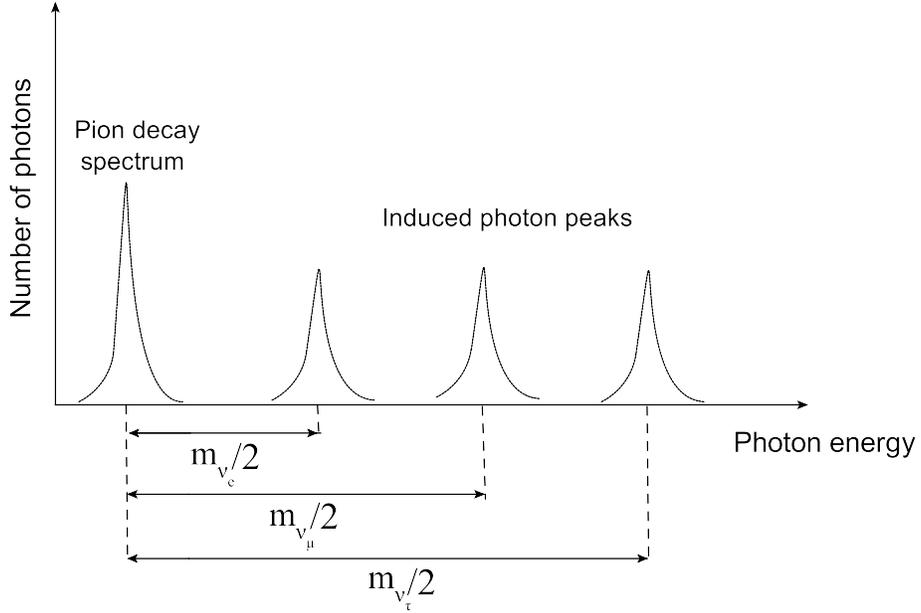}
}\caption{Qualitative illustration of the spectrum of photons emitted in the process $\pi^0\rightarrow~\gamma\gamma$ plus $\nu_{l}\pi^0\rightarrow\nu_l\gamma$. The neutrino mass hierarchy $m_{\nu_{e}}>m_{\nu_{\mu}}>m_{\nu_{\tau}}$ is assumed.}
\label{fig4}
\end{figure}

The process~(\ref{pion_d0}) is distinguishable from the decay

\begin{equation}
\pi^0\rightarrow\gamma+\gamma
\label{pion_d02}
\end{equation}

because of the non-zero neutrino mass. Again there will be a gap between the maximum energies of the photons emitted in~(\ref{pion_d0}) and~(\ref{pion_d02}) approximately equal to $m_{\nu_l}/2$. This situation is qualitatively illustrated in Fig.~\ref{fig4}.

Moreover, (\ref{pion_d0}) has a unique signature due to the emission of a single photon mimicking thus a single photon decay of $\pi^0$ (the outgoing neutrino will carry away half of the pion energy undetected simultaneously providing conservation of the angular momentum). So, observation of the single photons instead of the expected photon pairs in neutral pion decays will indicate interaction with the neutrinos.

Summing over the contributions of the three neutrino flavors (just multiplying~(\ref{cross_sect_tot2}) by 3) one finds the mean lifetime of $\pi^0$  with respect to~(\ref{pion_d0}) induced by any of the neutrinos

\begin{equation}
\bar\tau_{\pi}(\nu_{l})\approx1.27\times10^{28}\,\, \text{yr}\,\left(\frac{1.92\times10^{-45} \text{cm}^2}{3\sigma_{\pi}}\right)\left(\frac{56~\text{cm}^{-3}}{n_{\nu_{l}}}\right)\left(\frac{7.7\times10^{-4}c}{v}\right).
\label{pion0_life2}
\end{equation}

The CNB can manifest itself through~(\ref{pion_d0}) in matter with a pion condensate. Such a state of matter may be naturally formed in compact stars. The existence of these pions or, more generally, states with the quantum numbers of the pion may provide an additional neutrino (photon) emission channel apart from other possible ones~\cite{pion_cond1,pion_cond2,pion_cond3,pion_cond4}.

One can estimate the corresponding present-day neutrino (photon) emissivity of an object with the pion number density $n_{\pi}$:

\begin{equation}
\varepsilon_{\pi\nu(\gamma)}\approx8.5\times10^{11} \frac{\text{MeV}}{\text{cm}^3\,\text{yr}}\left(\frac{n_{\pi}}{1.6\times10^{38} \text{cm}^{-3}}\right)\left(\frac{1.27\times10^{28}\,\, \text{yr}}{\bar\tau_{\pi}(\nu_{l})}\right)\left(\frac{m_{\pi}/2}{67.5 \,\,\text{MeV}}\right).
\label{emiss}
\end{equation}

It is also interesting to consider possibilities of detection of the CNB in laboratory experiments which would exploit interaction of the CNB neutrinos with a pion condensate in atomic nuclei~(provided, of course, such nuclei exist).

\section{Induced charged pion decay}

The CNB neutrinos may also induce charged pion decays.  Consider the radiative decay

\begin{equation}
\pi^-\rightarrow l^-+\bar\nu_l+\gamma,
\label{pion_d}
\end{equation}

where $l=e, \mu$.

Crossing (\ref{pion_d}) yields

\begin{equation}
\nu_l+\pi^-\rightarrow l^-+\gamma,
\label{pion_d2}
\end{equation}

which is a non-threshold reaction as well.

\begin{figure}
\centering
\resizebox{1.0\textwidth}{!}{%
\includegraphics{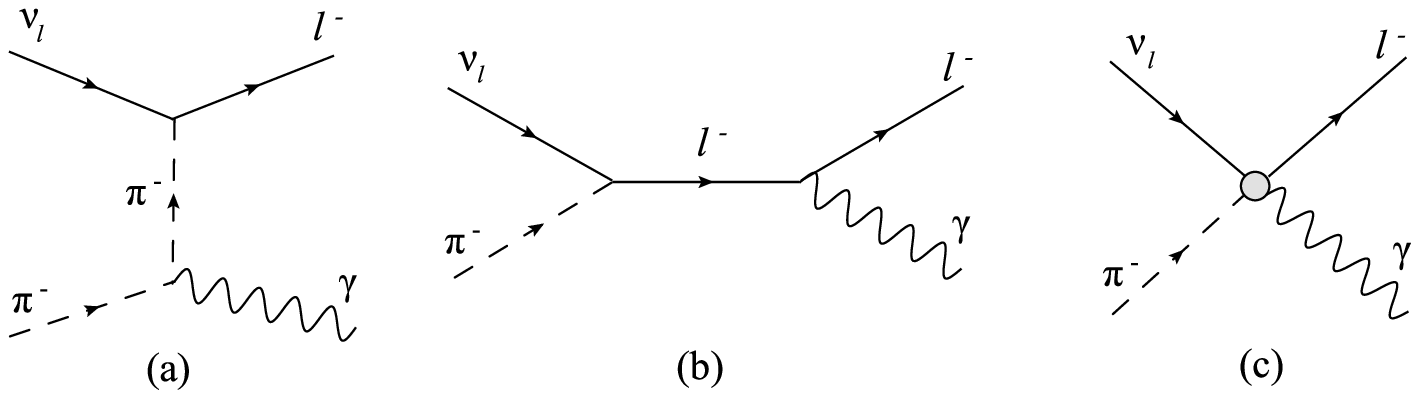}
}\caption{Feynman diagrams for the process $\nu_l\pi^-\rightarrow l^-\gamma$ ($l=e,\mu$).}
\label{fig5}
\end{figure}

The tree level Feynman diagrams contributing to (\ref{pion_d2}) are shown in Fig.~\ref{fig5}.

Let us consider only the incident electron neutrino. In this case the contributions of the diagrams (a) and (b) are helicity suppressed being proportional to $m_e$ exactly as in the case of the $\pi_{e2}$ decay. For this reason we  neglect them keeping only the diagram (c) which is free of the suppression. Then the matrix element takes the form~\cite{Chen2011}:

\begin{eqnarray}
{\cal M}=-i{\frac{G_{F}}{\sqrt{2}}}V_{ud}\varepsilon_{\mu }
\bar{u}(p_{e})\gamma _{\alpha }(1-\gamma
_{5})u(p_{\nu})\hskip 4cm\nonumber\\\times
\left[e\frac{F_{A}}{m_{\pi}}(-g^{\mu \alpha }p_{\pi}\cdot
q+p_{\pi}^{\mu }q^{\alpha })+ie\frac{F_{V}}{m_{\pi}}\epsilon ^{\mu \alpha \beta
\lambda }q_{\beta }p_{\pi \lambda }\right], \label{ampl_decay}
\end{eqnarray}

where $V_{ud}$ is the Cabibbo--Kobayashi--Maskawa matrix element, $F_{A}$ is
the axial-vector form factor. Throughout this Letter we set $F_V=0.0272$ and $F_A=0.0112$ \cite{Chen2011}.

The cross section of the reaction $\nu_e\pi^-\rightarrow e^-\gamma$ reads

\begin{equation}
\sigma^c_{\pi}=\frac{\alpha G^2_{F}m^2_\pi}{16}|V_{ud}|^2|F_{V}-F_{A}|^2 \label{cross_pic}
\end{equation}

and the corresponding mean lifetime is

\begin{equation}
\bar\tau^c_{\pi}(\nu_{e})\approx2.2\times10^{29}\,\, \text{yr}\,\left(\frac{1.06\times10^{-46} \text{cm}^2}{\sigma^c_{\pi}}\right)\left(\frac{56~\text{cm}^{-3}}{n_{\nu_{e}}}\right)\left(\frac{7.7\times10^{-4}c}{v}\right).
\label{pionc_life2}
\end{equation}

Details of similar calculations can be found in~\cite{plb_pion}.

The relation between the spectra of the emitted electrons (photons)  in (\ref{pion_d}) and (\ref{pion_d2}) will be analogous to the ones discussed above.

\section{Conclusions}
Several non-threshold reactions which may be used to detect the CNB neutrino background are presented. Namely $\nu_{\mu}\mu^+\rightarrow e^+\nu_e$, $\bar\nu_{e}\mu^+\rightarrow e^+\bar\nu_{\mu}$, $\nu_{l}\pi^0\rightarrow\nu_{l}\gamma$ and $\nu_{e}\pi^-\rightarrow e^-\gamma$.  The corresponding cross sections are calculated analytically within the Standard Model. It is notable that these reactions are sensitive not only to the electron neutrinos  but also to the muon and tau neutrino content of the CNB.

The CNB may manifest itself through these reactions in interactions with objects containing states with the quantum numbers of the pion (compact stars and/or atomic nuclei).

The presented analysis directly applies to other pseudoscalar mesons as well. For example, the cross sections and the other quantities for the kaon are obtainable just by performing the appropriate replacements in the formulae given in this Letter.

It is interesting that the discussed reactions mimic lepton number violation as well as angular momentum non-conservation. If we had not expected the existence of the CNB, we would have thought of these reactions in experiment as lepton number violating decay modes of $\mu$ and $\pi$.

\vskip 0.8cm
{\bf Acknowledgements}

This work was supported in part by the Russian Foundation for Basic Research (grant 11-02-12043), by the Program for Basic Research of the Presidium of the Russian Academy of Sciences "Fundamental Properties of Matter and Astrophysics" and by the Federal Target Program  of the Ministry of Education and Science of Russian Federation "Research and Development in Top Priority Spheres of Russian Scientific and Technological Complex for 2007-2013" (contract No. 16.518.11.7072).













\newpage

\begin{thebibliography}{00}
\bibitem{indirect}
S.~Hannestad, Ann. Rev. Nucl. Part. Sci. 56 (2006) 137.

\bibitem{revm1}
V. F. Shvartsman, V. B. Braginskii, S. S. Gershtein, Ya. B. Zel'dovich, M. Yu. Khlopov, JETP Lett. 36 (1982) 277.

\bibitem{revm2}
P. F. Smith, J. D. Lewin, Phys. Lett. B 127 (1983) 185.

\bibitem{ringwald09}
A. Ringwald, 	Nucl. Phys. A 827 (2009) 501c.

\bibitem{rev1}
S. Weinberg, Phys. Rev. 128 (1962) 1457.

\bibitem{rev2}
J. M. Irvine, R. Humphreys, J. Phys. G 9 (1983) 847.

\bibitem{rev3}
A. G. Cocco, G. Mangano, M. Messina, JCAP 0706 (2007) 015
[J. Phys. Conf. Ser. 110 (2008) 082014].


\bibitem{rev5}
M. Blennow, Phys. Rev. D 77 (2008) 113014.

\bibitem{rev6}
A. Kaboth, J. A. Formaggio, B. Monreal, Phys. Rev. D 82 (2010) 062001.

\bibitem{rev7}
Y. F. Li,  Zhi-zhong Xing, Shu Luo, Phys. Lett. B 692 (2010) 261.

\bibitem{rev4}
R. Hodak, S. Kovalenko, F. Simkovic, A. Faessler, arXiv:1102.1799.

\bibitem{mu2e}
F. Cervelli, J. Phys. Conf. Ser. 335 (2011) 012073.

\bibitem{cluster}
A. Ringwald, Y. Y. Y. Wong, JCAP 0412 (2004) 005.

\bibitem{comp_star}
A. Schmitt, Dense Matter in Compact Stars: A Pedagogical Introduction, Springer, Berlin and Heidelberg, 2010.

\bibitem{Harvey2007}
J. A. Harvey, Ch. T. Hill, R. J. Hill, Phys. Rev. Lett. 99 (2007) 261601.

\bibitem{plb_pion}
I. Alikhanov, Phys. Lett. B 706 (2012) 423.

\bibitem{pion_cond1}
P. Jaikumar, M. Prakash, T. Schaefer, Phys. Rev. D 66 (2002) 063003.

\bibitem{pion_cond2}

F. Arretche, A. A. Natale, D. N. Voskresensky, Phys. Rev. C 68 (2003) 035807.

\bibitem{pion_cond3}
S. Reddy, M. Sadzikowski, M. Tachibana, Nucl. Phys. A 714 (2003) 337.

\bibitem{pion_cond4}
M. Loewe, C. Villavicencio, arXiv:1107.3859.

\bibitem{Chen2011}
C.~H.~Chen, C.~Q.~Geng, C.~C.~Lih, Phys. Rev. D 83 (2011) 074001.



\end{thebibliography}
\end{document}